\documentclass[reprint,
superscriptaddress,
 amsmath,amssymb,
 aps,
pra,]{revtex4-1}
\usepackage{blindtext}
\usepackage{float}
\usepackage{graphicx}
\usepackage[colorlinks,urlcolor=blue,linkcolor=blue,anchorcolor=blue,citecolor=blue]{hyperref}
\usepackage{subfigure}
\begin{document}
\title{Strong field frustrated double ionization of argon atoms}
\author{Shi Chen}
\affiliation{School of Physics, Peking University, Beijing 100871, China}
\affiliation{Center for Applied Physics and Technology, HEDPS, and College of Engineering, Peking University, Beijing 100871, China}
\author{HuiPeng Kang}
\email{H.Kang@gsi.de}
\affiliation{Institute of Optics and Quantum Electronics, Friedrich Schiller University Jena, Max-Wien-Platz 1,07743 Jena, Germany}
\affiliation{Helmholtz Institut Jena, Frötbelstieg 3, 07743 Jena, Germany}
\affiliation{State Key Laboratory of Magnetic
Resonance and Atomic and Molecular Physics, Wuhan Institute of
Physics and Mathematics, Innovation Academy for Precision Measurement Science and Technology, Chinese Academy of Sciences, Wuhan 430071, China}
\author{Jing Chen}
\affiliation{Institute of Applied Physics and Computational Mathematics, P.O. Box 8009, Beijing 100088, China}
\affiliation{Center for Advanced Material Diagnostic Technology, College of Engineering Physics, Shenzhen Technology University, Shenzhen 518118, China}
\author{Gerhard G. Paulus}
\affiliation{Institute of Optics and Quantum Electronics, Friedrich Schiller University Jena, Max-Wien-Platz 1,07743 Jena, Germany}
\affiliation{Helmholtz Institut Jena, Frötbelstieg 3, 07743 Jena, Germany}

\begin{abstract}

Using a three-dimensional semiclassical method, we theoretically investigate frustrated double ionization (FDI) of Ar atoms subjected to strong laser fields. The double-hump photoelectron momentum distribution generated from FDI observed in a recent experiment [S. Larimian \textit{et al.}, Phys. Rev. Research 2, 013021 (2020)] is reproduced by our simulation. We confirm that the observed spectrum is due to recollision. The laser intensity dependence of FDI is investigated. We reveal that the doubly excited sates of Ar atoms and excited states of Ar$^{+}$ are the dominant pathways for producing FDI at relatively low and high intensities, respectively. Our work demonstrates that at modest intensities, FDI is a general strong-field physical process accompanied with nonsequential double ionization and it is an important consequence of recollision. 


\end{abstract}

\maketitle

\section{INTRODUCTION}
When exposed to a strong laser field, the outermost electron of atoms or molecules can be ionized through tunneling. The electron is then accelerated and possibly driven back by the oscillating laser electric field to recollide with its parent ion \cite{PhysRevLett.71.1994}, resulting in various strong-field phenomena such as above-threshold ionization (ATI) plateau \cite{becker2002above}, high-order harmonic generation (HHG) \cite{mcpherson1987studies,ferray1988multiple}, and nonsequential double ionization (NSDI) \cite{RevModPhys.84.1011}. Alternatively, due to the presence of the Coulomb field of the ion, the electron may also be captured into an Rydberg state after the end of the laser pulse, leading to excited atoms or molecules \cite{bing2006coulomb,PhysRevLett.101.233001,Shvetsov-Shilovski2009,PhysRevLett.109.093001,PhysRevLett.110.203002,PhysRevLett.114.123003,PhysRevA.93.033415,PhysRevA.94.053403}. This is known as frustrated tunneling ionization (FTI) \cite{PhysRevLett.101.233001}. The capture of electrons into Rydberg states was also found when strong field double ionization (DI) occurs \cite{Shomsky2009, PhysRevLett.102.113002}, which can be referred to frustrated double ionization (FDI). It has been mainly observed in atomic fragments produced by Coulomb explosion of molecules and also dimers \cite{PhysRevLett.102.113002,nubbemeyer2009excited,PhysRevA.82.013412,PhysRevA.82.013413,PhysRevA.84.043425,McKenna_2012}, which can be explained as neutralization during dissociative ionization process \cite{PhysRevLett.106.203001,PhysRevA.85.011402}. Taking FDI of hydrogen molecules for example \cite{PhysRevLett.102.113002,PhysRevA.85.011402}, two electrons tunnel ionize sequentially and one of them may be captured by one of the protons when the molecular ions fragment, leading to the formation of a highly excited neutral hydrogen atom and a proton. Experimentally, the capture process is identified by measuring the kinetic energies of the excited neutral fragments after molecular dissociative ionization. Unfortunately, this method is not applicable for atomic targets since atomic FDI happens without dissociation and the products are excited ions rather than excited neutral fragments. Consequently, there is a lack of experimental and theoretical studies on atomic FDI.   

Very recently, FDI of Ar atoms was experimentally identified by measuring the dc-field ionized electrons from the excited singly charged ions (Ar$^{+*}$), the photoelectrons, and the corresponding doubly charged ions in coincidence (Ar$^{2+}$) \cite{larimian2018frustrated}. The measured photoelectron momentum distributions corresponding to an FDI event display a clear transition from a double-hump to a single-hump structure as the laser intensity is increased, quite similar to DI. The observation suggests that the physical mechanism of FDI differs for different intensity regions where NSDI or sequential double ionization (SDI) dominate, respectively. For SDI occurring at high intensities, the two electrons are ionized independently. The narrow single-hump structure for FDI at such high intensities, which has a width close to that for single ionization, strongly suggests that the trapped electrons mainly arise from the second ionization step. For modest intensities, where NSDI dominates, it has been speculated that the electron-electron interaction during recollision causes the double-hump momentum spectrum for FDI \cite{larimian2018frustrated}. Yet, how exactly recollision results in such a spectrum remains unclear. 
 

In this paper, we theoretically study FDI of Ar atoms using a semiclassical model. The main purpose of the current work is to offer a transparent physical picture of FDI at modest intensities where NSDI dominates. We calculate the ratio of FDI to single ionization (SI) and the ratio of FDI to DI as functions of laser intensity. Our calculation reproduces the experimental double-hump photoelectron momentum distribution of FDI as reported in \cite{larimian2018frustrated}. We confirm that recollision is responsible for this structure and further show that how recollision leads to different photoelectron momentum distributions for FDI at different intensities. By analyzing the electron trajectories, we find that the dominant pathways for FDI at relatively low and high intensities are doubly excited states of Ar and excited states of Ar$^{+}$, respectively. This work indicates that FDI is a general strong-field process accompanied with NSDI and it is another important consequence of recollision.   

The paper is organized as follows. In Sec. \ref{sec.THEORETICAL MODEL}, we introduce the semiclassical model. Section \ref{sec.RESULTS AND DISCUSSION} shows our main results. Finally, we present our conclusion in Sec. \ref{CONCLUSION}.

\section{THEORETICAL MODEL}
\label{sec.THEORETICAL MODEL}
A well-established three-dimensional semiclassical model (see, e.g., \cite{PhysRevA.61.033402,PhysRevA.83.053422}) is employed to describe FDI and NSDI. In this model, we consider the interaction of a two-active-electron atom with a linearly polarized laser field: 
\begin{equation}
\mathbf{E}(t)=f(t) E_{0} \cos \omega t\ \hat{\mathbf{z}},
\end{equation}
where $\omega$ is the laser frequency and $E_0$ is the peak amplitude of the laser electric field. The pulse envelope function $f(t)$ is a constant equal to 1 for the first ten laser cycles and then reduced to zero with a three-cycle ramp in the form of $\cos^2$. 

Following the same procedure as used in previous studies \cite{chen2017non,PhysRevA.95.063415,li2017universal},  the outermost electron $e_1$ is assumed to be ionized by quantum tunneling through the field-suppressed atomic potential. The tunneling process can be described by the Schr\"odinger equation in parabolic coordinates (atomic units are used throughout this paper) \cite{landau2013quantum}:
\begin{equation}
\label{eqSE}
\frac{d^{2} \phi}{d \eta^{2}}+\left(\frac{I_{p 1}}{2}+\frac{1}{2 \eta}+\frac{1}{4 \eta^{2}}+\frac{E \eta}{4}\right) \phi=0,
\end{equation}
where $I_{p1}$ is the first ionization potential of atoms. Eq. (\ref{eqSE}) describes the tunneling process for an electron with energy of $I_{p1}/4$ within an effective potential $U(\eta)=-1 / 4 \eta-1 / 8 \eta^{2}-E \eta / 8$. Thus, the tunnel exit point $\eta_0$ is determined by solving the equation $U(\eta)=I_{p1}/4$. The corresponding initial positions of $e_1$ are $x_0=y_0=0$, $z_0=-\eta_0/2$. The initial longitudinal velocity is assumed to be zero and a nonzero initial velocity perpendicular to the laser polarization direction is introduced \cite{HUP1997533}. The corresponding initial velocities are thus $v_{x0}=v_\perp \cos \theta$, $v_{y0}=v_\perp \sin \theta$, and $v_{z0}=0$, where $\theta$ is the angle between the transverse velocity $v_\perp$ and the $x$-axis. For the bound electron $e_2$, its initial conditions are determined by assuming this electron in the ground state of singly charged ions and the corresponding positions and momenta are depicted by a microcanonical distribution \cite{PhysRevA.26.3008}. 
\begin{figure}
\centering
\includegraphics[width=1\columnwidth]{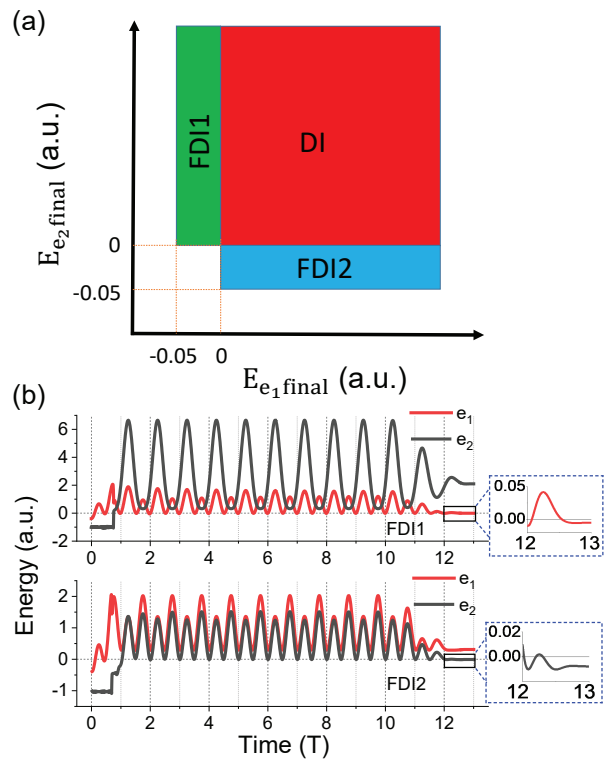}
\caption{\label{fig1} (a) Schematic diagram of different types of electron trajectories for Ar. $E_{e_1,\text{final}}$ and $E_{e_2,\text{final}}$ are the final energies of $e_1$ and $e_2$, respectively. See text for details. (b) Typical time evolutions (in laser cycle T) of the two-electron energies for FDI1 (upper panel) and FDI2 (lower panel).}
\end{figure}

After setting the initial conditions of $e_1$ and $e_2$, the propagation of the two electrons is governed by the classical Newtonian equation of motion:
\begin{equation}
\label{newton}
\frac{d^{2} \mathbf{r}_{\mathbf{i}}}{d t^{2}}=-\mathbf{E}(t)-\nabla\left(V_{n e}^{i}+V_{e e}\right)
\end{equation}
until the end of the laser pulse. The index i=1, 2 in Eq. (\ref{newton}) denotes $e_1$ and $e_2$, respectively. $V_{n e}^{i}=-\frac{2}{\left|\mathbf{r}_{i}\right|}$ and $V_{e e}=\frac{1}{\left|\mathbf{r}_{1}-\mathbf{r}_{2}\right|}$ are the Coulomb interactions between the nucleus and the \textit{i}th electron and between the two electrons, respectively.

In our model calculation, $10^8$ initial points are randomly distributed in the parameter space $-\pi/2<wt_0<\pi/2$, $v_\perp>0$ and $0<\theta<2\pi$ for $e_1$ and in the microcanonical distribution for $e_2$ at each laser intensity. Here $t_0$ is the tunneling ionization instant. The laser frequency $\omega$ is chosen as 0.05642 a.u., corresponding to laser wavelength of 800 nm. Each electron trajectory is weighted by $W\left(t_{0},\ v_{\perp}\right)=W_{0}\left(t_{0}\right) W_{1}\left(t_0,\ v_{\perp}\right)$, where
\begin{equation}
W_{0}\left(t_{0}\right)=I_{p1} C_{n^{*} l}^{2}\left[\frac{2\left(2 I_{p1}\right)^{\frac{3}{2}}}{|E(t_0)|}\right]^{2 n^{*}-1} \exp \left[-\frac{2\left(2 I_{p1}\right)^{\frac{3}{2}}}{3|E(t_0)|}\right]
\end{equation}
is the tunneling rate \cite{ammosov1986tunnel,Delone:91}. Here $C_{n^{*} l}=\left(\frac{2 e}{n^{*}}\right)^{n^{*}} \frac{1}{\sqrt{2 \pi n^{*}}}$ is a constant with the effective principal quantum number $n^{*}=\frac{1}{\sqrt{2 I_{p1}}}$ and the \textit{e} constant. $W_{1}\left(t_0,\ v_{\perp}\right)$ denotes the distribution of the transverse velocity $v_{\perp}$, which is given by
\begin{equation}
W_{1}\left(t_0,\ v_{\perp}\right)=\frac{v_\perp\left(2 I_{p1}\right)^{\frac{1}{2}}}{|E(t_0)| \pi} \exp \left[-\frac{v_{\perp}^{2}\left(2 I_{p1}\right)^{\frac{1}{2}}}{|E(t_0)|}\right].
\end{equation}

As shown in Fig. \ref{fig1}(a), different groups of electron trajectories can be identified, depending on the final energies of the two electrons. Here we choose Ar as the target. DI events are identified when the final energies of both electrons are larger than zero. FDI events are identified when one electron has positive final energy and the other is captured into highly excited Rydberg states $\mathrm{Ar^{+*}}$ after the end of the laser pulse. Depending on which electron is recaptured, FDI events can be distinguished into FDI1 events: $E_{e_2,\text{final}}>0>E_{e_1,\text{final}}>-0.05\ \mathrm{a.u.}$ and FDI2 events: $E_{e_1,\text{final}}>0>E_{e_2,\text{final}}>-0.05\ \mathrm{a.u.}$, respectively. The corresponding quantum number \textit{n} of $\mathrm{Ar^{+*}}$ is larger than 6 (the energy of $\mathrm{Ar^{+*}}$ is $\mathrm{E_{Ar^{+*}}}=-2/n^2$). We note that the analysis and main conclusion presented in this paper also hold true if we choose even higher quantum numbers (e.g., $n>$10 or 20). 

Figure \ref{fig1}(b) displays typical time evolutions of the two-electron energies for FDI1 and FDI2. In our calculation, the returning electrons do not directly populate the high-lying Rydberg states of $\mathrm{Ar^{+}}$ associated with FDI via impact excitation. Instead, both the energies of the two electrons can be larger than zero after recollision and one of them is captured at the end of the laser pulse, as shown in Fig. \ref{fig1}(b). 

\section{RESULTS AND DISCUSSION}
\label{sec.RESULTS AND DISCUSSION}

\begin{figure}
\centering
\includegraphics[width=1\columnwidth]{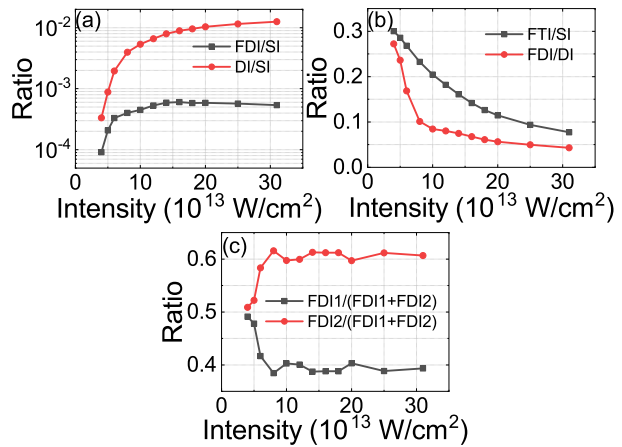}
\caption{\label{fig2}(a) Calculated ratios of FDI to SI and DI to SI as functions of laser intensity. (b) Same as (a) but for the ratios of FTI to SI and FDI to DI. (c) Calculated ratio of FDI1 to FDI (the sum of FDI1 and FDI2) and ratio of FDI2 to FDI as functions of intensity.}
\end{figure}

\begin{figure}
\centering
\includegraphics[width=1\columnwidth]{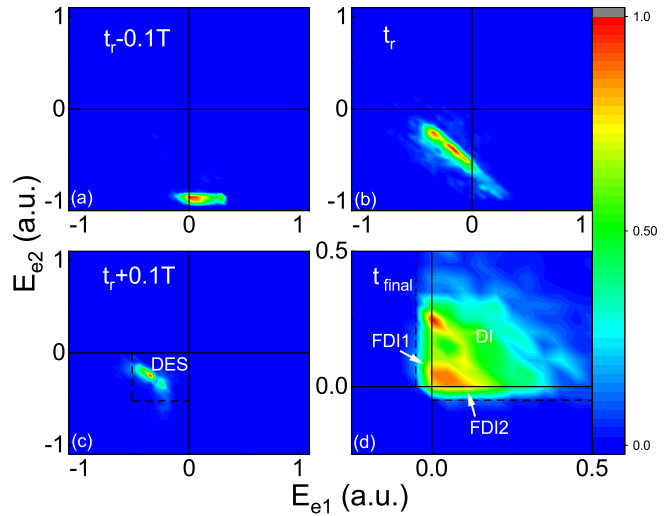}
\caption{\label{fig3}Electron-electron energy distributions at different times for the intensity of $4\times 10^{13}\ \mathrm{W} / \mathrm{cm}^{2}$. The two dashed lines in (c) for $\mathrm{E_{e_1}}=-0.52\ a.u.$ and $\mathrm{E_{e_2}}=-0.52\ a.u.$ are used to confine the region for doubly excited states of $\mathrm{Ar}$. Here $-0.52\ \mathrm{a.u.}$ is the energy of the first excited state of $\mathrm{Ar}$. The two dashed lines in (d) for $\mathrm{E_{e_1}}=-0.05\ a.u.$ and $\mathrm{E_{e_2}}=-0.05\ a.u.$ are plotted to confine the regions for FDI1 and FDI2, respectively. The color scale of each panel has been normalized for comparison purposes.}
\end{figure}

Figure \ref{fig2}(a) shows the calculated ratio of FDI to SI and the ratio of DI to SI as functions of laser intensity. The intensity-dependent ratio of DI to SI exhibits a pronounced knee structure, which is a characteristic feature of NSDI \cite{PhysRevA.27.2503,PhysRevLett.69.2642,PhysRevLett.73.1227,PhysRevA.52.R917,Talebpour_1997}. Interestingly, the ratio of FDI to SI shows a similar dependence on the intensity. The intensity-dependent ratio of FDI to DI is shown in Fig. \ref{fig2}(b). For comparison, we also calculate FTI events, which are identified when $E_{e_1,\text{final}}<0$ and $E_{e_2,\text{final}}=-I_{p2}$, where $I_{p2}$ is the second ionization potential of Ar. We find that both the calculated ratios of FTI to SI and FDI to DI decrease with the increase of intensity. Most FTI events are contributed by directly ionized trajectories rather than recollision trajectories because recollision tends to increase the drift momentum of $e_1$ \cite{Shvetsov-Shilovski2009,zhao2019frustrated}. As the intensity is increased, the directly ionized $e_1$ obtains larger momentum and the distance between $e_1$ and the ionic core becomes larger. Correspondingly, its kinetic energy becomes larger and the Coulomb attraction between $e_1$ and the core becomes weaker. As a result, it is harder for $e_1$ to be captured and the ratio of FTI to SI decreases as the intensity is increased. In our calculation, the ratio of FTI to SI decreases from 0.3 to 0.077 when the intensity is increased from $4\times 10^{13}\ \mathrm{W/cm^2}$ to $3.1\times 10^{14}\ \mathrm{W/cm^2}$. This is in excellent agreement with the theoretical derivation in Ref. \cite{Shvetsov-Shilovski2009} that this ratio will decrease from 0.3 to 0.072 ($\mathrm{Ar^*/Ar^+}\propto \frac{1}{I^{3/4}}(1-\frac{\sqrt{I}}{2I_{p1}^2})^{-1}$, where $I$ is the laser intensity) for the same range of intensities. As for FDI, all the events are due to recollision in our calculation. When the intensity is increased, the returning energy of $e_1$ becomes larger and both the energies of $e_1$ and $e_2$ right after recollision thus become larger. Hence, it is harder for any one of the two electrons to be captured at the end of the laser pulse and the ratio of FDI to DI decreases.

Figure \ref{fig2}(c) shows the ratios of FDI1 to FDI and FDI2 to FDI as functions of laser intensity. In the relatively high intensity regime ($I>7\times 10^{13}\ \mathrm{W} / \mathrm{cm}^{2}$), the probability of FDI2 is significantly higher than the probability of FDI1. In the low intensity regime ($I\leq 5\times 10^{13}\ \mathrm{W} / \mathrm{cm}^{2}$), the probability of FDI1 is close to that of FDI2. For even lower intensities, the probability of FDI is extremely low and it is very challenging to calculate.


In order to understand the dependence of FDI on intensity, we investigate the energy
distributions of the two electrons at $4\times 10^{13}\ \mathrm{W} / \mathrm{cm}^{2}$ and $3.1\times 10^{14}\ \mathrm{W} / \mathrm{cm}^{2}$, respectively. Figures \ref{fig3}(a)-\ref{fig3}(c) display the electron-electron energy distributions of both DI and FDI events around the recollision time (denoted as $t_r$) when the two electrons are closest to each other for $4\times 10^{13}\ \mathrm{W} / \mathrm{cm}^{2}$. Right before $t_r$, the returning energy of $e_1$ has a cutoff of 0.33 a.u., which is slightly larger than 3.17$U_p$ ($U_p$ is the ponderomotive energy) due to the existence of the tunneling exit and the Coulomb potential. The energy of $e_2$ is around -1.02 a.u. ($-I_{p2}$). At $t_r$, $e_1$ transfers some energy to $e_2$. We find that doubly excited states (DESs) of $\mathrm{Ar}$ are largely populated shortly after $t_r$ and the binding energy of $e_1$ is close to that of $e_2$ [Fig. \ref{fig3}(c)], which is consistent with previous studies of NSDI \cite{PhysRevLett.112.013003}. As $e_1$ and $e_2$ share the energy evenly during recollision and experience the same laser electric field afterwards, one would expect no preference for each electron with higher final energy than the other one. Hence, the probabilities of FDI1 and FDI2 are close to each other, as shown in Fig. \ref{fig2}(c). 


\begin{figure}
\includegraphics[width=1\columnwidth]{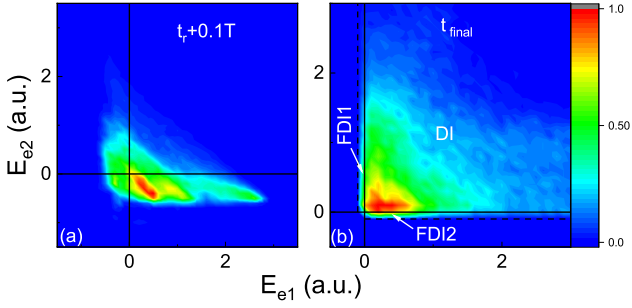}
\caption{\label{fig4} (a) and (b) Electron-electron energy distributions at different times for the intensity of $3.1\times 10^{14}\ \mathrm{W} / \mathrm{cm}^{2}$. The dashed lines have the same meanings as Fig. \ref{fig3}(d). The color scales have been normalized for
comparison purposes.}
\end{figure}

\begin{figure}
\centering
\includegraphics[width=1\columnwidth]{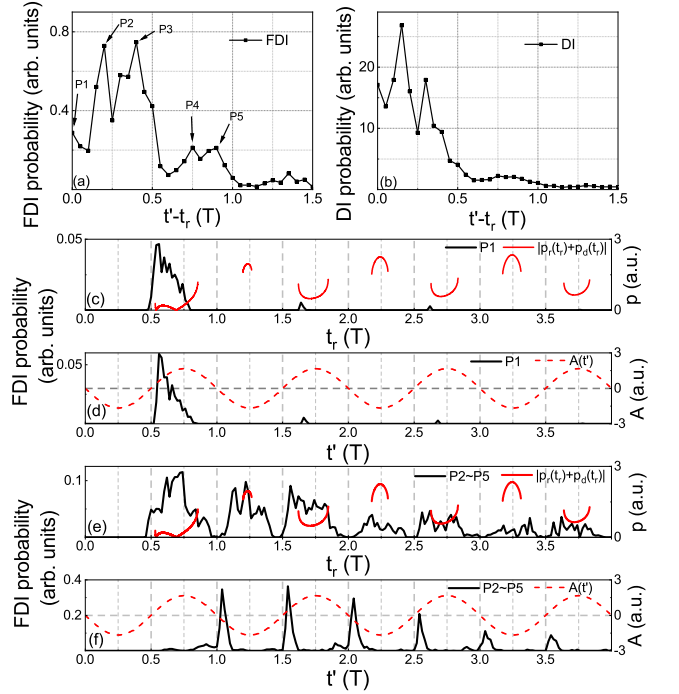}
\caption{(a) and (b) Time delay distributions for FDI and DI events at the intensity of $3.1 \times 10^{14}\ \mathrm{W/cm^2}$, respectively. (c) and (d) Probability distributions (black curves) of $t_r$ and $t^\prime$ for FDI events corresponding to P1 in (a), respectively. Corresponding distributions of $\left|p_r(t_r)+p_d(t_r)\right|$ (red curves) and $A(t^\prime)$ (red dotted lines) are also shown. See text for details. (e) and (f) Same as (c) and (d) but for FDI events corresponding to P2$\sim$P5 in (a), respectively.}
\label{fig5}
\end{figure}

As seen in Fig. \ref{fig4}(a), the energy of $e_1$ shortly after recollision is larger than zero for most DI and FDI events at $3.1\times 10^{14}\ \mathrm{W} / \mathrm{cm}^{2}$, which indicates that the DESs are no longer the main pathways for producing DI and FDI events for such high intensity. At $3.1\times 10^{14}\ \mathrm{W} / \mathrm{cm}^{2}$, NSDI of Ar proceeds mainly via recollision impact ionization (RII) \cite{PhysRevLett.87.043003}. Despite the formation of $\mathrm{Ar^{+*}}$, the excited electron $e_2$ is ionized quickly after recollision. It has been recently shown that, both theoretically \cite{li2017universal} and experimentally \cite{kang2018steering}, there exists a time delay lasting for a small fraction of T between $t_r$ and double ionization time for RII. Here we show that a significant time delay between $t_r$ and $t^\prime$ occurs for both FDI and DI events [Figs. \ref{fig5}(a) and \ref{fig5}(b)], where $t^\prime$ is the instant when both the energies of $e_1$ and $e_2$ are larger than zero for the first time for FDI events [see Fig. \ref{fig1}(b)] and the double ionization time for DI events. In accordance with Ref. \cite{li2017universal}, the time delay distribution shows three pronounced peaks for DI events [Fig. \ref{fig5}(b)]. As for FDI events, our calculation reveals five pronounced peaks, denoted as P1$\sim$P5 in Fig. \ref{fig5}(a). In the following, we will show that this time delay distribution is the key to understand the ratio of FDI1(FDI2) to FDI at high intensities shown in Fig. \ref{fig2}(c).

Firstly, we discuss the mechanism of FDI events corresponding to the first peak (P1) in Fig. \ref{fig5}(a). The sum energy of the two electrons right after recollision is $E_{r_1}-I_{p2}$, where $E_{r_1}$ is the returning energy of $e_1$. We have found that the energy of $e_2$ right after recollision is close to zero for most FDI events corresponding to P1. Neglecting the Coulomb potential effect, the final energies of the two electrons can thus be expressed as 
\begin{equation}
\label{Ee1}
E_{e_1,\text{final}}\approx\frac{[p_r(t_r)+p_d(t_r)]^2}{2}=\frac{[\sqrt{2(E_{r_1}-I_{p2})}-A(t_r)]^2}{2}
\end{equation} 
and
\begin{equation}
\label{Ee2}
E_{e_2,\text{final}}\approx\frac{p_d(t^\prime)^2}{2}=\frac{A(t^\prime)^2}{2}\ ,
\end{equation}
respectively, where $p_r(t_r)$ is the residual momentum of $e_1$ right after recollision and $p_d$ is the drift momentum obtained from the laser field subsequently. In Fig. \ref{fig5}(c) we show the probability distribution of $t_r$ for P1 and corresponding distribution of the sum of $p_r(t_r)$ and $p_d(t_r)$, predicted by the simple-man theory \cite{paulus1994rescattering}. One can find that $t_r$ is mainly distributed from 0.5T to 0.75T where the sum of $p_r(t_r)$ and $p_d(t_r)$ is close to zero, indicating that $p_r(t_r)$ is cancelled out by $p_d(t_r)$. According to Eq. (\ref{Ee1}), $E_{e_1,\text{final}}$ is thus close to zero. As shown in Fig. \ref{fig5}(d), the probability distribution of $t^\prime$ is similar to that of $t_r$ and the vector potential $A(t^\prime)$ is nonzero for most trajectories. Consequently, for P1, $e_1$ is more likely to be captured after the end of the laser pulse, leading to FDI1. 

For most FDI events corresponding to P2$\sim$P5 in Fig. \ref{fig5}(a), the energy of $e_1$ is larger than zero while the energy of $e_2$ is slightly smaller than zero right after $t_r$, which is similar to Fig. \ref{fig4}(a). As shown in Fig. \ref{fig5}(e), $p_r(t_r)$ and $p_d(t_r)$ do not cancel each other out for multiple returning trajectories, which contribute to FDI significantly. Consequently, the final energy of $e_1$ can be quite large for most FDI events. As for $e_2$, it stays in the excited state of $\mathrm{Ar^+}$ until $t^\prime$. Figure \ref{fig5}(f) displays the corresponding probability distribution of $t^\prime$. One can find that the distribution peaks around $\frac{n+1}{2}T\ (n=1,2,3,\dots)$ where the vector potential $A(t^\prime)$ is close to zero. This is different from the case for P1. Therefore, $e_2$ tends to be captured after the end of the laser pulse [Eq. (7)], leading to FDI2. Due to much more contribution of P2$\sim$P5 as compared with P1 [Fig. \ref{fig5}(a)], the probability of FDI2 is larger than that of FDI1, as shown in Fig. \ref{fig2}(c).

\begin{figure}
\includegraphics[width=1\columnwidth]{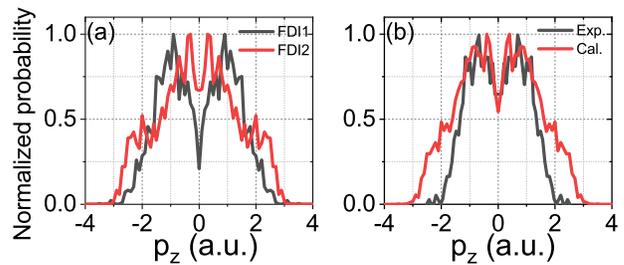}
\caption{\label{fig6} (a) Calculated photoelectron momentum distributions for FDI1 and FDI2 at $3.1\times 10^{14}\ \mathrm{W/cm^2}$. (b) Comparison between the calculated photoelectron momentum distribution for FDI at $3.1\times 10^{14}\ \mathrm{W/cm^2}$ and corresponding experimental result in Ref. \cite{larimian2018frustrated}.}
\end{figure}

To compare with the experiment, we further calculate the photoelectron momentum distribution for FDI at $3.1 \times 10^{14}\ \mathrm{W/cm^2}$. As shown in Fig. \ref{fig6}(a), the photoelectron momentum distribution for FDI1 displays a double-hump structure. For the calculation of FDI2, two additional ``shoulders" around $p_z=\pm $2 a.u. can be seen. For FDI2, the photoelectron ($e_1$) has the final momentum approximately equal to $p_r(t_r)+p_d(t_r)$, which is close to zero for the first-return-collision trajectories [Fig. \ref{fig5}(e)], leading to the shallow dip around $p_z=$ 0 a.u. shown in Fig. \ref{fig6}(a). The sum of $p_r(t_r)$ and $p_d(t_r)$ is much larger for even-order-return-collision trajectories [Fig. \ref{fig5}(e)], leading to the shoulder-like structures around $p_z=\pm $2 a.u. For FDI1, $e_1$ is captured at the end of the laser pulse. The momentum of $e_2$ right after recollision is close to zero. As shown in Fig. \ref{fig5}(d), the distribution of $t^\prime$ peaks around 0.6T and the corresponding vector potential $A(t^\prime)$ is 0.97 a.u. [$A(t^\prime)=-\frac{E}{w}\sin (wt^\prime)$]. Therefore, the final momentum distribution of the photoelectrons peaks around $\pm$0.97 a.u. [Eq. (7)], which is consistent with the position of the double-hump structure for FDI1 in Fig. \ref{fig6}(a). 


Figure \ref{fig6}(b) compares the calculated photoelectron momentum distribution for FDI (the sum of FDI1 and FDI2) and the experimental result in Ref. \cite{larimian2018frustrated}. The calculation is in good agreement with the experiment, confirming above analysis based on the recollision picture. The calculated probabilities for $|p_z|>2$ a.u. are higher than the experiment. This can be partially attributed to the intensity volume effect in the experiment, which is not included in our calculation. 

\begin{figure}
\centering
\includegraphics[width=1\columnwidth]{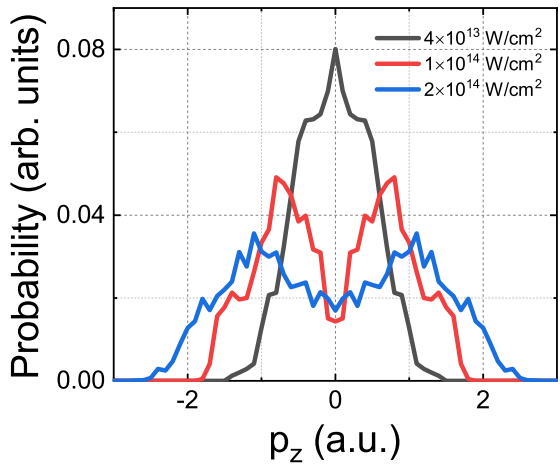}
\caption{\label{fig7} Calculated photoelectron momentum distributions for FDI at different intensities.}
\end{figure}

Finally, we show the calculated photoelectron momentum distributions for FDI at different intensities in Fig. \ref{fig7}. The intensity dependence can be understood as follows. For FDI2 events,  with the decrease of the intensity, $p_r(t_r)$ becomes smaller than $p_d(t_r)$ so that the final momenta of the photoelectrons ($e_1$) are no longer close to zero. This suppresses the FDI2 events around $p_z=$ 0 a.u. As a result, the dip of the electron momentum distribution for FDI becomes more pronounced. When the intensity is further decreased to the regime where the DESs of Ar are the dominant pathways leading to FDI ($I\leq 5\times 10^{13}\ \mathrm{W} / \mathrm{cm}^{2}$), the two electrons stay in the DESs for a while and one of them is then ionized around the maximum of the laser field where the vector potential is close to zero. Therefore, the final momenta of the photoelectrons peak around zero, leading to a single-hump distribution with a much narrower width. The predicted intensity-dependent photoelectron momentum distributions can be testified by further experiments. 

\section{CONCLUSION}
\label{CONCLUSION}
In conclusion, we theoretically investigate intensity dependence of FDI process of Ar atoms. The calculated ratio of FDI to SI as a function of intensity shows a pronounced knee structure and the ratio of FDI to DI decreases with the increase of intensity. In the relatively low intensity regime, we demonstrate that the DESs of Ar are the dominant pathways for producing FDI. The probabilities of FDI1 and FDI2 are close to each other and the photoelectron momentum distribution shows a single-hump structure. For the relatively high intensity regime, the excited states of Ar$^{+}$ are the dominant pathways leading to FDI. The probability of FDI2 is significantly higher than that of FDI1 and the photoelectron momentum distribution exhibits the double-hump structure, which is in good agreement with the recent experimental result \cite{larimian2018frustrated}. Our work confirms that this observation is due to recollision and explains that how recollision results in different photoelectron momentum distributions for different intensities. This work demonstrates that FDI generally exists as a companion with strong-field NSDI process and offers intuitive physical insights into FDI of atoms.   


\section{ACKNOWLEDGEMENT}
\label{ACKNOWLEDGEMENT}
We appreciate valuable discussions with YanLan Wang, YueMing Zhou and XinHua Xie. This work is supported by the National Key Research and Development Program of China (No. 2019YFA0307700 and No. 2016YFA0401100), the National Natural Science Foundation of China (No. 11974380), and the German Science Foundation (PA 730/6). 
\bibliographystyle{apsrev4-1} 
\bibliography{FDI} 
\end{document}